\documentclass[preprint]{aastex63}
\usepackage{natbib}

\graphicspath{{./}{figures/}}

\usepackage{hyperref}
\received{}
\revised{}
\accepted{}
\submitjournal{ApJ}

\shorttitle{Turbulence and heating}
\shortauthors{Yordanova et al.}
\graphicspath{{./}{figures/}}

\begin{document}

\title{A possible link between turbulence and plasma heating}


\correspondingauthor{Emiliya Yordanova}
\email{eya@irfu.se}

\author[0000-0002-9707-3147]{Emiliya Yordanova}
\affiliation{Swedish Institute of Space Physics, Uppsala, Sweden}

\author[0000-0001-7597-238X]{Zolt\'{a}n V\"{o}r\"{o}s}
\affiliation{Space Research Institute, Austrian Academy of Sciences, Graz, Austria}

\affiliation{Institute of Earth Physics and Space Science, E\"otv\"os Lor\'and Research Network, Sopron, Hungary}

\author[0000-0002-5981-7758]{Luca Sorriso-Valvo}
\affiliation{Swedish Institute of Space Physics, Uppsala, Sweden}

\affiliation{CNR - Istituto per la Scienza e Tecnologia dei Plasmi, Bari, Italy.}

\author{Andrew P. Dimmock}
\affiliation{Swedish Institute of Space Physics, Uppsala, Sweden}


\author{Emilia Kilpua}
\affiliation{Department of Physics, University of Helsinki, Helsinki, Finland.}

\begin{abstract}


Numerical simulations and experimental results have shown that current sheets formation in space plasmas can be associated with enhanced vorticity. Also, in simulations the generation of such structures is associated with strong plasma heating. 
Here, we compare four-point  measurements in the terrestrial magnetosheath turbulence from the Multiscale magnetospheric mission (MMS) of the flow vorticity and the magnetic field curlometer versus their corresponding one-point proxies PVI(V) and PVI(B) based on the Partial Variance of Increments (PVI) method. 
We show that the one-point proxies are sufficiently precise in identifying not only the generic features of the current sheets and vortices statistically, but also their appearance in groups associated with plasma heating. 
The method has been further applied to the region of the turbulent sheath of an interplanetary coronal mass ejection (ICME) observed at L1 by WIND spacecraft. 
We observe current sheets and vorticity associated heating in larger groups (blobs), which so far have not been considered in the literature on turbulent data analysis. The blobs represent extended spatial regions of activity with enhanced regional correlations between the occurrence of conditioned currents and vorticity, which at the same time are also correlated with enhanced temperatures. This heating mechanism is substantially different from the plasma heating in the vicinity of the ICME shock, where plasma beta is strongly fluctuating and there is no vorticity. The proposed method describes a new pathway for linking the plasma heating and plasma turbulence, and it is relevant to in-situ observations when only single spacecraft measurements are available.

\end{abstract}

\keywords{discontinuities, current sheets, turbulence, plasma heating, coronal mass ejections}

\section{Introduction} \label{sec:int}
Heating in astrophysical plasmas in absence of collisions still remains an unsolved question. Already the first spacecraft missions revealed that the solar corona is much hotter than the solar surface. Even more remarkably, the solar wind during its expansion does not cool adiabatically, implying that local mechanisms are continuing to heat the particles in the interplanetary space. 
Although MHD models of turbulence transport can predict, for example, the profiles of fluctuating kinetic energy, density and proton temperature with heliocentric distance, the turbulent model heating depends on  parametrization and phenomenological assumptions without a direct incorporation of local microphysics of kinetic energy dissipation  (\citealp[]{Zank17}, \citealp[]{Adhikari20}).  To understand local heating mechanisms in collisionless plasmas the corresponding kinetic plasma processes have to be understood.
To date, there are several decades of in-situ spacecraft observations of the solar wind in the heliosphere and detailed measurements in the near Earth's plasma regions that evidenced one or another physical mechanism of collisionless plasma heating. These include ---to name a few--- linear wave damping, stochastic heating \citep{Chandran10}, magnetic reconnection, microinstabilities \citep[and references therein]{Gary15} and plasma turbulence \citep[and references therein]{Chen19}. However, their relative contribution and importance are still largely unknown. Reconnecting, turbulence generated small-scale (sub-ion scale) current sheets are also considered as a possible physical mechanism associated with small-scale energy conversion and dissipation. Recent simulations show that the intermittently occurring current sheets in turbulence are accompanied by other coherent structures, such as vortices, density structures, gradients of different quantities \citep{Matthaeus21}, and at kinetic level, by distortions of the pressure tensor \citep{DelSarto16}. These coherent structures are not overlapping, but spatially separated over distances comparable to ion inertial length ($d_i$). Recent analysis based on both fluid and kinetic scale plasma simulations and by using multispecies Vlasov-Maxwell formulations suggest that the physical quantities associated with multi-scale energy transfer and energy conversions show the same type of intermittent and localized spatial concentrations nearby the coherent structures and gradients \citep{Matthaeus20}.
This indicates that there might exist several channels of energy conversion/dissipation and plasma heating associated with spatially concentrated small-scale coherent structures \citep{Matthaeus21}.
There is an on-going debate whether the plasma heating is concentrated at thin current sheets (with a thickness of the order  $d_i$) (\citealp[]{Pezzi18}, \citealp[]{Valentini16}), nearby the current sheets at plasma vortices \citep{Jain17}, or at locations of enhanced pressure-strain interactions \citep{Chasapis18a}. The latter is associated with the energy conversion between flow and random particle motions, therefore in fully kinetic simulations the pressure-strain work was found to be correlated with velocity gradients \citep{Yang17}.  Although there are no local pointwise correlations between the coherent structures and physical quantities associated with energy transfer and conversion, these are co-located within regions of size comparable to $d_i$. Therefore, some coarse-grained or regional correlations can be expected between them \citep{Yang19}. The generation of intermittent small-scale coherent structures is theoretically not fully understood. It might be the consequence of self-organization processes operating as the turbulent cascade is transferring energy from large scales towards the kinetic scales \citep{Matthaeus15}. Similarly, the explanation of the  occurrence of magnetically dominated meso-scale structures, such as flux ropes, magnetic islands, their interactions and relation to the turbulent energy transfer rate represent further theoretical challenges. Nevertheless, in the solar wind, magnetic islands and current sheets represent the basic ingredients in explaining plasma heating and particle acceleration. 
Small-scale islands can also result from magnetic reconnection within the heliospheric current sheet (HCS) \citep{Zhao20b} or various interactions between large-scale structures, such as the HCS and stream interfaces \citep{Adhikari19}, and the HCS and the heliospheric termination shock \citep{Zhao19}. Local acceleration observed at 5 AU was attributed to the interaction of such magnetic islands \citep{Zhao18}.

In this work, we focus specifically on the heating in turbulent plasmas that is associated with the generation of specific structures by turbulence. Observations in the solar wind have revealed that current layers arising from turbulence develop complex sub-proton networks of secondary islands and very thin currents sheets in the regions of reconnection outflows \citep{Greco16}. In simulations, these thin current sheets have been associated with strong electron heating \citep{Hesse01}. We aim to use proxies for vorticity and current sheets that can be calculated from one-point measurements in the solar wind, where also the time resolution of field and plasma measurements is limited. As a consequence, the identification of thin ion-electron scale structures potentially associated with plasma heating might not be resolved. Our working hypothesis is that due to turbulent intermittency, heating can happen in larger scale blobs, which are presumably much larger than the ion inertial length. In fact, the meaning of turbulent intermittency is that the turbulent energy is spatially non-homogeneous, i.e. there are voids in space with less kinetic energy and volumes where the energy is concentrated. In those blobs, the occurrence of vortical flows which might potentially generate current sheets is presumably also enhanced.  
In such 'active' blobs there might exist an "integrated" elevated temperature, relative to the ambient plasma, which could be observed more easily than the heating signatures at localized ion/electron scale structures. Also, the limited time resolution of single spacecraft in the solar wind would be less restrictive for the identification of larger-scale active blobs. 
However, the usefulness of one-point proxies has to be tested. To this end, we use multi-point measurement techniques for physical quantities such us vorticity and current density from the MMS mission. This allows us to test the four-point high time resolution estimation of vorticity  $\mid\nabla\times\textbf{\textit{V}}\mid$ and curlometer $\mid\nabla\times\textbf{\textit{B}}\mid$ versus one-point proxies PVI(V) and PVI(B) with smaller temporal resolution. Although we do not expect one-to-one correspondence, the proxies should allow us to correlate the turbulence-generated intermittent structures with intermittent heating in the solar wind. In order to test the proxies on high quality estimators of $\mid\nabla\times\textbf{\textit{V}}\mid$ and $\mid\nabla\times\textbf{\textit{B}}\mid$, we have selected intervals of MMS observations in the turbulent terrestrial  magnetosheath. After showing that the proxies work in the magnetosheath we apply the same approach to the turbulent sheath region of an interplanetary coronal mass ejection (ICME). This way we attempt to understand the integrated role of groups of thin current sheets and the associated vorticity in active intermittent blobs to global plasma heating rather than the heating at individual current sheets.

\section{Observations and method} \label{sec:dat}

For the analysis in this study, we use in-situ data from the multipoint Multiscale Magnetospheric (MMS) mission and the solar wind monitor WIND, positioned at L1. The MMS observations were performed on the day-side terrestrial magnetosheath (MSH) proper and the flanks (shown in Figure \ref{fig:fig1}), while the WIND measurements originate from an event of relatively fast ICME.

\subsection{MMS observations}

The four MMS spacecraft are equipped with the same state-of-the-art suite of instruments providing unprecedented high temporal and spatial resolution plasma and electromagnetic field measurements \citep{Burch16}. The spacecraft form a tetrahedron configuration at very close separation ($\sim$ tens of km). The main scientific objective of the mission is to resolve magnetic reconnection and kinetic scales plasma processes. We use magnetic field data measured by the FGM instrument \citep{Russell16} at sampling frequency 128 Hz; and electron and ion moments from the FPI instrument \citep{Pollock16} at sampling rate 150 and 30 ms, respectively. 

\begin{figure}[ht!]
 \centering
 \includegraphics[width=0.6\linewidth]{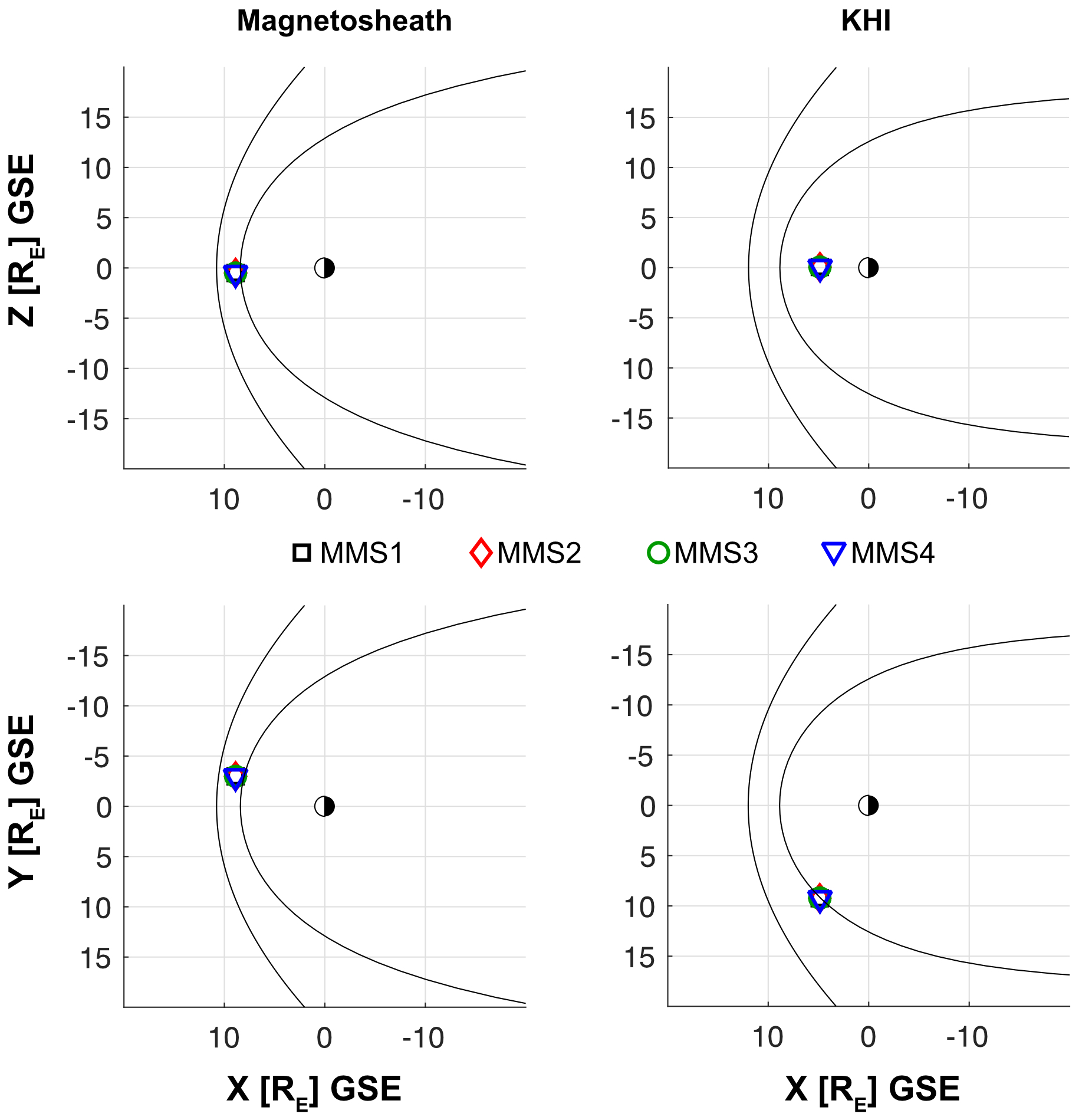}
\caption{MMS position in XZ and XY planes in GSE frame during the magnetosheath (MSH)  (\textit{left}) and during the Kelvin-Helmholtz  instability (KHI) intervals (\textit{right}). The inner parabolic curve represents the modeled magnetopause and the outer one - the bow shock using 1h upstream data from OMNI database. Note that MMS  separation is much smaller than the Earth radius $ R_E$ therefore the spacecraft symbols are overlapping.}
\label{fig:fig1}
\end{figure}

Two separate time intervals of MMS multipoint observations are considered. The first time interval consists of MMS observations of the turbulent MSH on 30 Nov 2015 between 00:21-00:26 UTC. During this time the MSH was downstream a quasi-parallel bow shock (Fig.\ref{fig:fig1}, \textit{left panels}). Thanks to the recent multipoint Cluster \citep{Escoubet01} and MMS missions, it has been well established that the MSH under this geometry is characterized by strong and intensive fluctuations and it is very dynamic \citep[and references therein]{Echim21}. Observations also indicate that under such turbulent conditions current sheets of the characteristic scale of the ion inertial length are present at high rates of occurrence (\citealp[]{Voros16}, \citealp[]{Yordanova20}). At turbulent current sheets various kinetic processes take place, such as plasma heating  (\citealp[]{Chasapis15}, \citealp[]{Chasapis17}, \citealp[]{Chasapis18b}), plasma acceleration \citep{Erik16}, magnetic reconnection (\citealp[]{Phan07}, \citealp[]{Retino07}, \citealp[]{Yordanova16}, \citealp[]{Voros17}, \citealp[]{Phan18}) and turbulent dissipation (\citealp[]{Sundkvist07}, \citealp[]{Voros19b}).

The second interval is an observation of Kelvin-Helmholtz (KH) vortices at the dusk-side magnetopause on 8 Sep 2015, between 10:07-11:25 UTC (Fig.\ref{fig:fig1}, \textit{right panels}). This period has been studied previously in terms of spectral and scaling properties of turbulence (\citealp[]{Stawarz16}, \citealp[]{Quijia21}). In another study \cite{Sorriso19} in the same KH interval, a strong connection was found between the local turbulent energy transfer at the end of inertial-range and the development of non-Maxwellian features of the ion Velocity Distribution Functions (VDFs), such as parallel ion beams. In addition, plasma jets at reconnecting current sheets were reported in the same event \citep{Eriksson16}.

\subsection{WIND observations}

We also use in-situ measurements from L1 from WIND spacecraft during an ICME passage on $17-18$ March 2013. This ICME was associated with a significant solar energetic particle (SEP) event and triggered a strong geomagnetic storm \citep{wu13}. We analyze 3 s solar wind proton data from the WIND 3-D Plasma and Energetic Particle Investigation instrument \citep{lin95}, and 0.092 s magnetic field observations from the Magnetic Field Investigation (MFI) instrument \citep{lepping95}. We take particular interest in the sheath region of the ICME. The ICME sheaths are formed when a CME ejecta propagates sufficiently faster than the ambient interplanetary plasma (\citealp[]{Kilpua17}, \citealp[]{Kilpua19}), producing a shock wave, behind which the solar wind plasma is collected and compressed by the expanding CME. Recent studies of the ICME sheaths (\citealp[]{Kilpua20}, \citealp[]{Kilpua21}) show that they are turbulent regions, with fluctuations amplitude, compressibility, and intermittency levels higher than that of the background solar wind. 

Even though the three cases considered here, namely a turbulent magnetosheath, a dusk-side magnetopause KH interval and an ICME sheath, may seemingly represent different plasma regions, they share in common a high level of turbulence, and that the plasma is compressed and bounded by boundaries. The velocity shear at the magnetopause leads to a growth of KH instability rolling-in layers of plasma with magnetosheath and magnetospheric origin in vortices. In the MSH, the plasma is confined between the bow shock and the magnetopause, while the ICME sheath is confined by the interplanetary (IP) shock and the leading edge of the ICME magnetic ejecta. It is worth noting, however, that the MSH and ICME sheath have different origins \citep{Siscoe08}. The former is of a ``propagation" type of a sheath layer, where the incoming solar wind flows sideways the obstacle (Earth) from the nose of the bow shock, while the latter is in general a combination of a ``propagation" and ``expansion" type. As the CME expands, the flow is deflected around it but at much slower speed resulting in a plasma pile-up instead of lateral flows around.

\subsection{Methods}

The Partial Variance of Increments (PVI) method has been broadly used in the analysis of turbulent plasmas (\citealp[]{Greco09}, \citealp[]{Matthaeus15}, \citealp[and references therein]{Greco18};  \citealp[]{Yordanova20}, \citealp[]{Kilpua20}) since its first application in MHD simulations (\citealp[]{Greco08}, \citealp[]{Servi11}). The PVI is a tool for identification of discontinuities in the magnetic field and plasma parameters. In single-point observations, the PVI is calculated by estimating the magnetic field increments, that is the magnitude of the vectorial differences $\Delta \textbf{\textit{Q}}(t,\tau)= \textbf{\textit{Q}}(t) - \textbf{\textit{Q}}(t+\tau$) of a quantity \textbf{\textit{Q}}, which is then normalized by the square root of their variance:

\begin{equation}
PVI(t,\tau)=\sqrt{\frac{\mid{\Delta \textbf{\textit{Q}}(t,\tau)\mid}^2}{\big \langle{\mid{\Delta \textbf{\textit{Q}}(t,\tau)\mid ^2}\big \rangle}}},
\label{eq1}
\end{equation}
where the averaging is done over the entire data sample and $\tau$ is the time delay between two instances of measurement, which defines the scale of the fluctuations of interest. Assuming the validity of the Taylor hypothesis \citep{Taylor38}, which supposes that the inherent plasma structures do not evolve in time or evolve much slower than the plasma flows past the spacecraft, the temporal scales can be transformed into spatial scales. The PVI can be calculated from a vector field, e.g. velocity or magnetic field, or a scalar, such as plasma density. 
The availability of multipoint Cluster and MMS measurements allowed for the method to be adapted and applied to two-point measurements from pairs of spacecraft (\citealp[]{Chasapis15}, \citealp[]{Voros16}, \citealp[]{Yordanova16}). In that case, Eq.~\ref{eq1}, is rewritten as:

\begin{equation}
PVI_{ij}(t)=\sqrt{\frac{\mid{\Delta \textbf{\textit{Q}}_{ij}(t)\mid}^2}{\big \langle{\mid{\Delta \textbf{\textit{Q}}_{ij}}\mid ^2}\big \rangle}},
\label{eq2}
\end{equation}
where again the average $\langle \cdot \rangle$ is taken over the whole data interval, and $i,j=$1,2,3,4 refers to the spacecraft number. The increments $\Delta \textbf{\textit{Q}}_{ij}(t)= \textbf{\textit{Q}}_i(t) - \textbf{\textit{Q}}_j(t)$ are now estimated between two points of measurement (i.e. pairs of spacecraft),
thus they represent the typical fluctuations over the actual spatial distance between the spacecraft. This means that PVI from multi-point measurements is sensitive to structures with sizes comparable to the distance between spacecraft.

In similar fashion, one can calculate the rotation or shear angles in a given vector field, from single point measurements:

\begin{equation}
\alpha(t,\tau)=\cos^{-1}\frac{\textbf{\textit{Q}}(t)\cdot \textbf{\textit{Q}}(t,\tau)}{\mid\textbf{\textit{Q}}(t)\mid \cdot \mid\textbf{\textit{Q}}(t,\tau)\mid},
\label{eq3}
\end{equation}

and from multi-point measurements:
 
\begin{equation}
\alpha_{ij}(t)=\cos^{-1}\frac{\textbf{\textit{Q}}_i(t)\cdot \textbf{\textit{Q}}_j(t)}{\mid\textbf{\textit{Q}}_i(t)\mid \cdot \mid\textbf{\textit{Q}}_j(t)\mid}.
\label{eq4}
\end{equation}

The calculation of the shear angle is useful in complementing the PVI method to distinguish between discontinuities (PVI$>$3) with or without large rotations, indicating the presence or absence of current sheets \citep{Chasapis15}.

An additional advantage of multi-point measurements is the estimation of the spatial gradient tensors across the tetrahedron configuration of spacecraft. We can calculate the gradient tensor of the magnetic field and the vorticity of the velocity field as $\nabla \times \textbf{\textit{Q}}$, where $\textbf{\textit{Q}}=\textbf{\textit{B}},\textbf{\textit{V}}$
\citep{Pashmann08}. 
In general, the turbulence-generated coherent structures at scales near the end of the fluid cascade are associated with small-scale gradients, also reflecting the abundance of available turbulent energy there, which can potentially generate various structures, i.e. current sheets.

\section{Results} \label{sec:res}

\begin{figure}[ht!]
 \centering
 \includegraphics[width=1\linewidth]{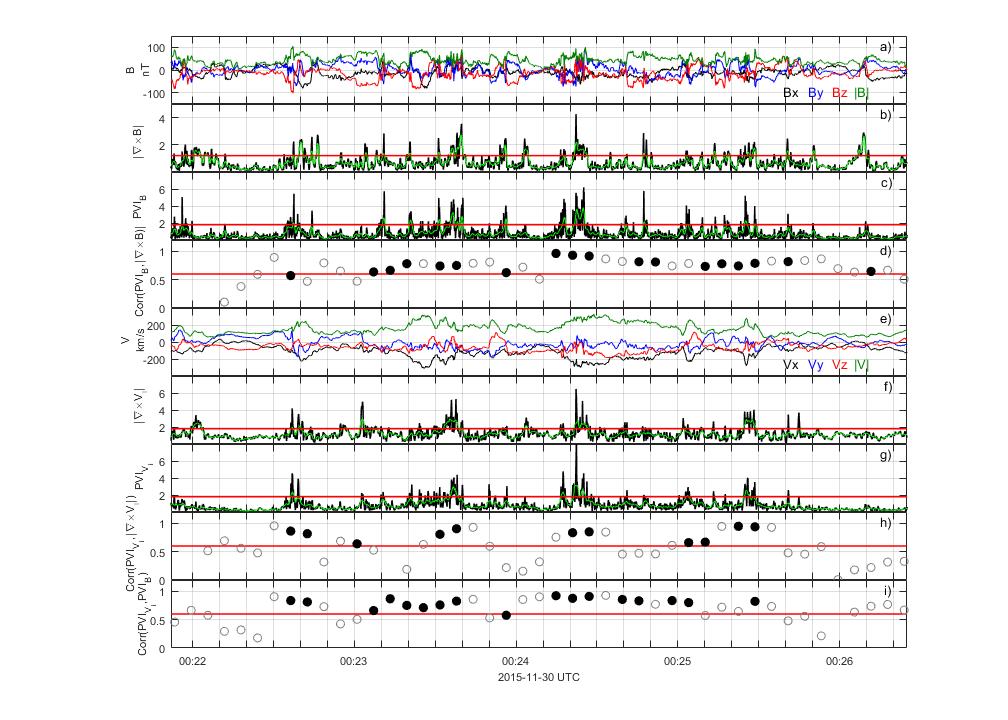}
\caption{Magnetosheath observations by MMS. \textit{(top to bottom)}: $a$) magnetic field components and magnitude for MMS~1 in GSE frame; $b$) curlometer from the magnetic field from all spacecraft; $c$) PVI(B) for MMS~1; $d$) Correlation between PVI(B) and $\mid\nabla \times \textbf{\textit{B}}\mid$; $e$) ion velocity components and magnitude in GSE frame for MMS1; $f$) ion vorticity from all spacecraft; $g$) PVI(V$_{i}$) for MMS~1; $h$) Correlation between PVI(V$_i$) and ion vorticity; $i$) Correlation between PVI(V$_i$) and PVI(B). The black circles in the correlations panels mark the correlation coefficients $cc$ coming from the quantities above their respective thresholds for $cc\geq$ 0.6; the empty gray circles are correlations from noise, i.e. from values under the thresholds. The green curves in $b,c,f,g$ represent the smoothed parameters used in the correlations calculation.}
\label{fig:fig2}
\end{figure}

\begin{figure}[ht!]
 \centering
 \includegraphics[width=1\linewidth]{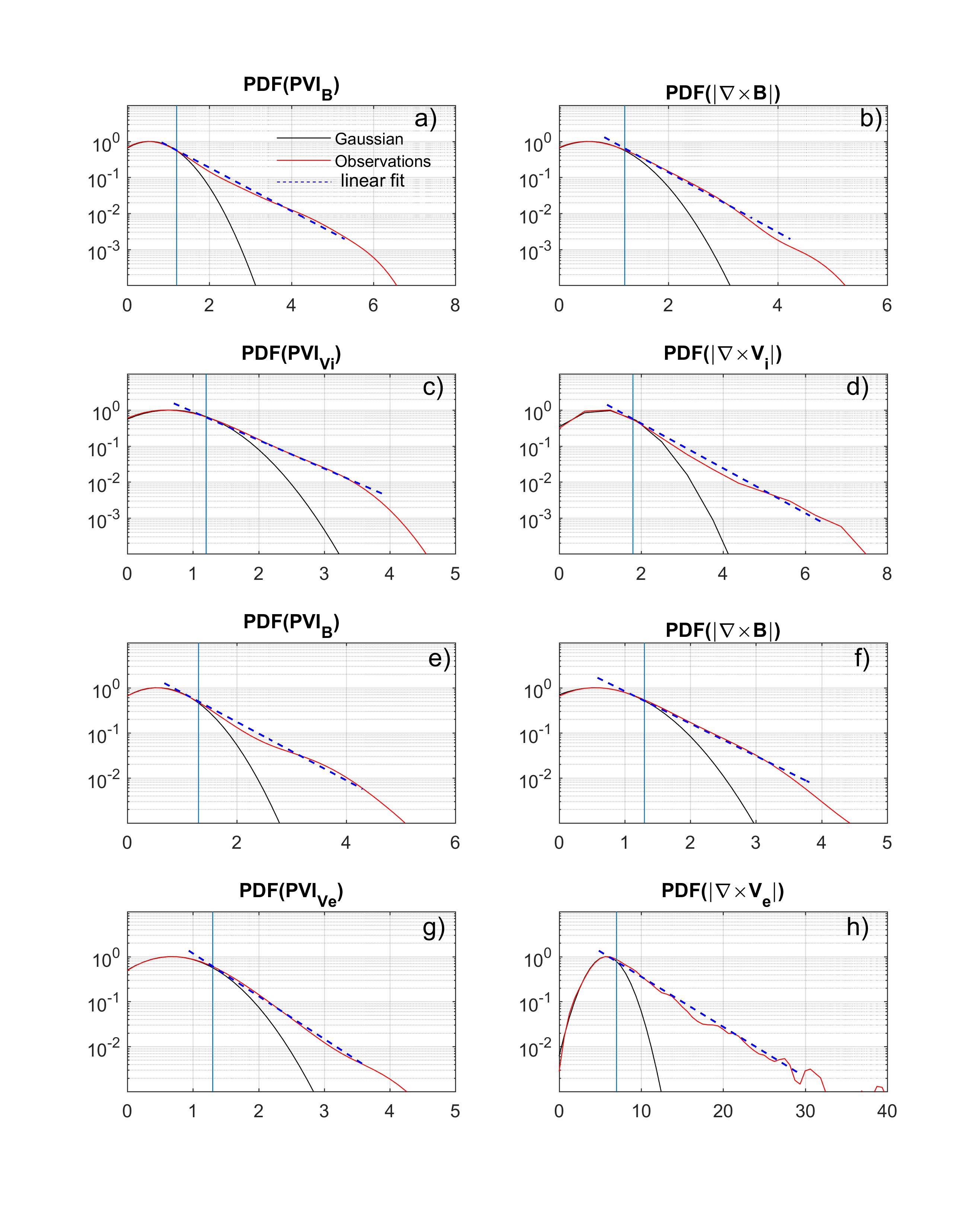}
\caption{PDFs (in red) of the smoothed  PVI(B), PVI(V), $\mid\nabla \times \textbf{\textit{B}}\mid$ in units [$mA/m^2$], and $\mid\nabla \times \textbf{\textit{V}}\mid$ in [$1/s$]. In black are shown the respective Gaussian distributions. 
The vertical light blue lines mark the thresholds where the deviations from the Gaussian are observed. Above these thresholds the long PDF tails can be fitted linearly (blue dashed lines). Panels $a-d$ ($e-h$) represent the quantities related to ion (electron) observations. The respective smoothing procedure is described in the text.}
\label{fig:fig3}
\end{figure}

\begin{figure}[ht!]
 \centering
 \includegraphics[width=1\linewidth]{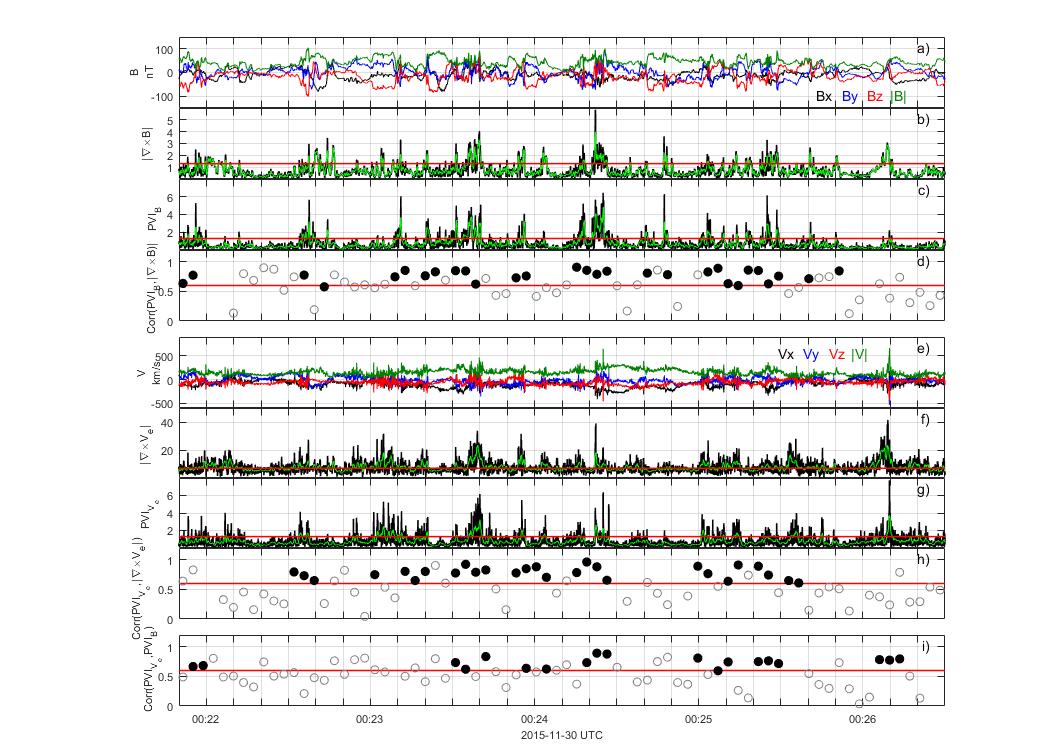}
\caption{Magnetosheath observation by MMS. \textit{(top to bottom)}: $a$) magnetic field components and magnitude in GSE frame; $b$) curlometer from the magnetic field from all spacecraft; $c$) PVI(B) for MMS~1; $d$) Correlation between PVI(B) and $\mid\nabla \times \textbf{\textit{B}}\mid$; $e$) electron velocity components and magnitude for MMS~1 in GSE frame; $f$) electron vorticity from all spacecraft; $g$) PVI(V$_{e}$) for MMS~1; $h$) Correlation between PVI(V$_e$) and electron vorticity; $i$) Correlation between PVI(V$_e$) and PVI(B). The colored circles in the correlations panels mark the correlation coefficients $cc$ coming from the quantities above their respective thresholds for $cc\geq$ 0.6; the empty gray circles are correlations from noise, i.e. from values under the thresholds. The green curves in $b,c,f,g$ represent the smoothed parameters used in the correlations calculation.}
\label{fig:fig4}
\end{figure}

Figure \ref{fig:fig2} shows MMS observations of $\sim 4$ min interval in the turbulent quasi-parallel magnetosheath. We compare results obtained by single-point and multi-point techniques. The top panel $a)$ shows the fluctuations of the magnetic field. In panel $b)$ the absolute values of the curlometer $\mid\nabla \times \textbf{\textit{B}}\mid$ \citep{Dunlop02} (in black) from the magnetic field are plotted, calculated from the four MMS spacecraft, and resampled to match the ion velocity. Overlaid with green is the smoothed curve at $\sim$1.65 s of $\mid\nabla \times \textbf{\textit{B}}\mid$. We have chosen a smoothing corresponding to the ion inertial length scale ($\sim$ 24 km here), because the coherent structures around a current sheet are not fully overlapping, but separated over distances comparable to $\sim d_i$. In our case, the smoothing for the ions is equal to 247 km which for a plasma flow of 150 km/s corresponds to $\sim$ 10 $d_i$. Such coarse graining by smoothing is done to reach better correlations between currents and vortices. The red horizontal line marks the threshold equal to 1.2, above which the fluctuations deviate from a Gaussian distribution. Fig. \ref{fig:fig3} (panels $a-d$) represent the PDF distributions of the smoothed ion quantities (red) compared to the respective Gaussian distributions (black). The thresholds are shown as the vertical light blue lines and are determined empirically as the value at which the observed PDF develop a power law tail (linear fit as blue dashed lines) significantly deviating from the Gaussian. In order to check for dependence of the threshold, the analysis was repeated using different values, giving similar results.
In Fig. \ref{fig:fig2}, panel $c)$ the PVI from MMS-1 (in black) is plotted, obtained for the time delay $\tau \sim $0.15 s, which is close to the delay of 0.12 s, corresponding to the separation distance in a spacecraft pair of $\sim$ 18 km for average bulk flow of $\sim$150 km/s. 
The PVI(B) time series of the other spacecraft look very similar because the spacecraft are close to each other. Again, the same smoothing technique is applied and the horizontal red line depicts the threshold of 1.2 (see also Fig. \ref{fig:fig3}, $a$). The physical meaning of PVI conditioning has been justified in simulations \citep{Greco08} and confirmed in observations \citep{Voros16}, and enables us to infer the enhanced current densities to non-Gaussian PVI(B). From Fig. \ref{fig:fig2}, panel $d)$ we can see that there is a good correlation ($>0.6$) between the PVI(B) and $\mid\nabla \times \textbf{\textit{B}}\mid$ at the times where both quantities are over the thresholds. We use Pearson correlation to measure  their linear dependence. The correlation coefficient is defined as: $cc=\frac{1}{N-1}\sum_{i=1}^{N}(\frac{A-\mu_A}{\sigma_A})(\frac{B-\mu_B}{\sigma_B})$, where $N$ is the number of data points, and $\mu$ and $\sigma$ denote the mean and standard deviations of the quantities $A$ and $B$. The correlation coefficient is calculated with running average in 18.5 s windows with $75\%$ overlapping. The black circles depict the correlation coefficients $cc\geq$ 0.6 that come from PVIs and rotations in the magnetic field both being above their determined thresholds, while the gray empty circles come from one or both of the correlating quantities being under the thresholds. These are kept in order to retain a reasonable number of points when calculating the correlations. The same annotation applies for the correlations in panels $h), i)$ and the respective panels in Figure \ref{fig:fig4}.
Similarly, in Fig. \ref{fig:fig2}, panels $e)-g)$ show the velocity field, the absolute value of the ion vorticity from four spacecraft, the corresponding PVI(V$_i$) for MMS$-1$ and their correlation with the thresholds obtained the same way. 
The correlation between PVI(V$_i$) and the vorticity is also evident in the case of the ion velocity during the time intervals where the clusters of peaks in both parameters well exceed the thresholds, namely 1.2 for PVI(V$_i$) and 1.8 for the vorticity (see, Fig. \ref{fig:fig3}, $c, d$). 
The correlation between single point PVI(V$_i$) and the magnitude of the multi-point vorticity is an unexpected result, demonstrating that this important quantity associated with the turbulent kinetic energy of the flow can perhaps be estimated with good approximation using single point in-situ measurements in the solar wind.
The last panel $i)$ of Fig. \ref{fig:fig2}, demonstrates the correlations between PVI(V$_i$) and PVI(B), where we can see that there are blobs of activity in both quantities. This seems to support our working hypothesis that the intermittent excess of the kinetic energy of turbulent plasma manifested by the enhanced vorticity or flow shears is associated with the occurrence of magnetic coherent structures in spatial blobs. Moreover, single-point measurements can be used to find this important correspondence between the ion velocity and magnetic field fluctuations.

Figure \ref{fig:fig4} follows the logic of Figure \ref{fig:fig2}, presenting the results for the same MSH interval only with the difference that we plot here the quantities related to the electrons - electron velocity V$_e$ (panel $e$), electron vorticity (panel $f$), PVI(V$_e$) (panel $g$), correlations between PVI(V$_e$) and electron vorticity (panel $h$), and between PVI(V$_e$) and PVI(B) (panel $i$).  The time delay is 0.12 s corresponding again to the spacecraft separation, and the windows length for the correlation averages and the smoothing (in green)  is of the order of $\sim$ 3.5 $d_i$, so that we see better the details in correlations.  
The red lines in the panels mark the non-Gaussian thresholds in PVI (panels $c,g$), and between B and V shears (panels $b,f$), as well as correlations thresholds in panels $d),h),i)$.
The parameters based on electron velocity in Figure \ref{fig:fig4} show similar characteristics as those based on ion velocity in Figure \ref{fig:fig2}, namely simultaneous deviations from Gaussianity with values 1.3 for $\mid\nabla \times \textbf{\textit{B}}\mid$, PVI(B) and  PVI(V), and 7 for $\mid\nabla \times \textbf{\textit{V}}\mid$ (Fig. \ref{fig:fig3}, $e-h$). However, due to the higher electrons sampling rate, there are finer details that are well visible in the correlation in Fig. \ref{fig:fig4}, panels $h)-i)$ and also there are some different features, which will be discussed in the next section.

\begin{figure}[ht!]
 \centering
 \includegraphics[width=1\linewidth]{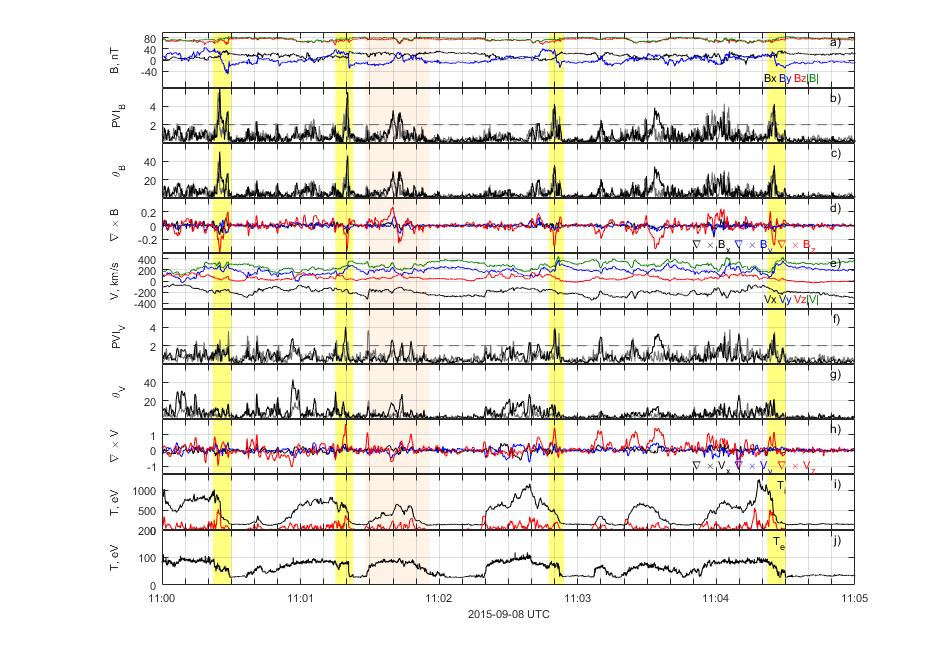}
 \caption{MMS observation of KH vortices. \textit{(top to bottom)}: $a$) magnetic field magnitude and components for MMS4; $b$) single-point PVI(B) (gray) from MMS4 and two-point PVI(B) calculated for MMS2 and MMS4 pair of spacecraft with time shift $\tau=0.6$ corresponding to the spacecraft separation; $c$) magnetic shear angle from MMS4 (gray), and from MMS2 and MMS4 (black); $d$) $\mid\nabla \times \textbf{\textit{B}}\mid$ components from all spacecraft; $e$) velocity magnitude and components for MMS4; $f$) PVI(V) for the pair MMS2-4 (black) overlaid on PVI(V) for MMS4; $g$) velocity shear angle for MMS4 (gray), and the pair MMS2 and MMS4 (black); $h$) ion vorticity components from all spacecraft; $i$) ion temperature - measured (black) and filtered (red) for MMS4; and $j$) electron temperature for MMS4. The horizontal dashed gray lines mark the PVI threshold. The shaded intervals highlighting current sheets are discussed in detail in the text.}
\label{fig:fig5}
\end{figure}

Next, we present the testing results for the KH interval (Fig. \ref{fig:fig5}). In general, KH vortices are large-scale structures reaching the size of several R$_E$ at the flank of the magnetosphere. In the considered event of 8 Sep 2015, 10:07-11:25 UTC, the average size of vortices was $\sim$ 2.8 R$_E$. The MMS spacecraft were positioned at $\sim (4.9, 9.1, 0.1) R_E$ in GSE frame, at a separation of $\sim$ 175 km, which in the time domain corresponds to 0.6 s for average plasma bulk 275 km/s. The entire interval is about two hours long. For the sake of better visualization of details, a 5-minute long time interval is shown in the figure. The KH vortices can be easily identified as the alternation of all parameters as the MMS spacecraft cross in succession the different plasma regions. 
From previous Cluster  observations with larger spacecraft separations (thousands of km) \citep{Hasegawa04}, it was shown that due to the relative plasma motion between the fast MSH flow at the flanks of the magnetopause and the stagnant magnetospheric low latitude boundary layer (LLBL) plasma, KH instability (KHI) can grow. When KHI has grown sufficiently the high-density plasma from MSH and the low density LLBL plasma on both sides of the magnetopause get engulfed and en-rolled into vortices bearing the characteristics of the two plasma regions. 
If we focus on the interval 11:02:00-11:02:50, the part that belongs to the MSH plasma is characterized by a stronger magnetic field (panel $a$), higher ion velocity (panel $e$), and lower ion and electron temperatures (panel $i, j$), while the part dominated by the LLBL has reverse features - lower magnetic field and velocity, and higher temperature. Another characteristic feature is the presence of periodic current sheets formed in the process of the vortex enrolling (in color shades). For example, the current sheets at $\sim$ 11:01:20 and $\sim$ 11:02:50 are detected by both single and multi-spacecraft techniques: strong peaks in the PVI(B) (panel $b$), where the gray curve is the single-point measurement and the black - the two-point one; together with the magnetic field shear angle (Eqs. 3, 4) (panel $c$, same color nomenclature) coming from the sharp changes in the magnetic field direction (panel $a$) observed by MMS 4, which are also detected at the same times with the spacecraft tetrahedron and appear as large peaks in the component of $\mid\nabla\times\textbf{\textit{B}}\mid$ along $Z$, which is the direction of vortex in-rolling (panel $d$). We have also calculated the single-point and two-point PVI(V$_i$) (panel $f$), ion velocity shear angle (Eqs. 3, 4) (panel $h$, in gray and black, respectively), and the three components of the four-point ion vorticity (panel $h$). The vortices' detection can be easily seen by following the flow rotation and velocity gradients in these panels.
Characteristic for the KH instability, is also that the temperatures are periodically changing because of plasma mixing (panel $i, j$). To see if there are local peaks in the temperature that are not intrinsic to the different regions but might be associated with enhanced PVI(B, V), rotation angles, and vorticity, we filter out the large scale variation in the ion temperature due to different regions crossings (red curve in panel $i$). We applied the same technique for the electron temperature but the structures were not very well seen.
Indeed, the filtered T$_i$ reveals local peaks that seem to occur in the intervals of the CSs (in color shades). 

\begin{figure}[ht!]
 \centering
 \includegraphics[width=0.5\linewidth]{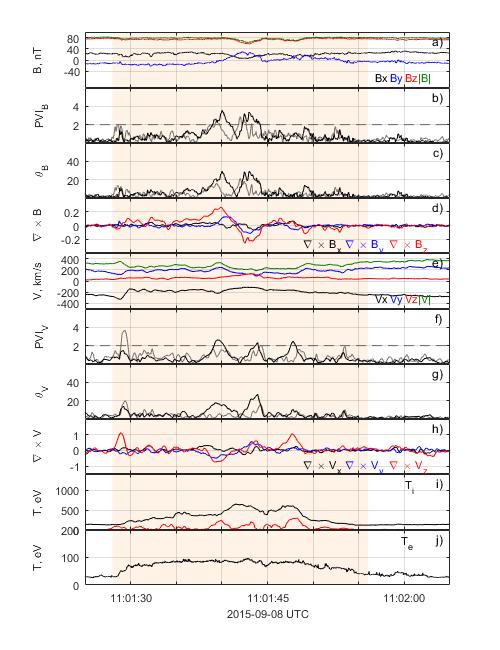}
  \caption{The interval 11:01:28-11:01:56 from Fig.  \textbf{\ref{fig:fig5}} zoomed in to highlight in the shaded area - the enhanced ion temperature (in red, panel $i$), associated with a group of current sheets (panels $b-d$) and large PVI(V$_i$), velocity shear and vorticity components (panel $f-g$).}
 \label{fig:fig6}
\end{figure}

In Fig. \ref{fig:fig6} we exemplify this by zooming in into the orange shaded interval. Outside the shaded box, the plasma velocity and the magnetic field are very quiet and the ion temperature is steady and non-fluctuating. Inside the highlighted interval, however, there is enhanced activity - at the times of the peaks in PVI(B) and magnetic field rotation (panels $b-d$), there are also strong velocity shears and vorticity (panels $f-h$). Remarkably, the local increases in the ion temperature correlated quite precisely with the increases in the discussed parameters, which strongly suggests that the observed ion heating is associated with the currents sheets generated by the enhanced vorticity.

\begin{figure}[ht!]
 \centering
 \includegraphics[width=1\linewidth]{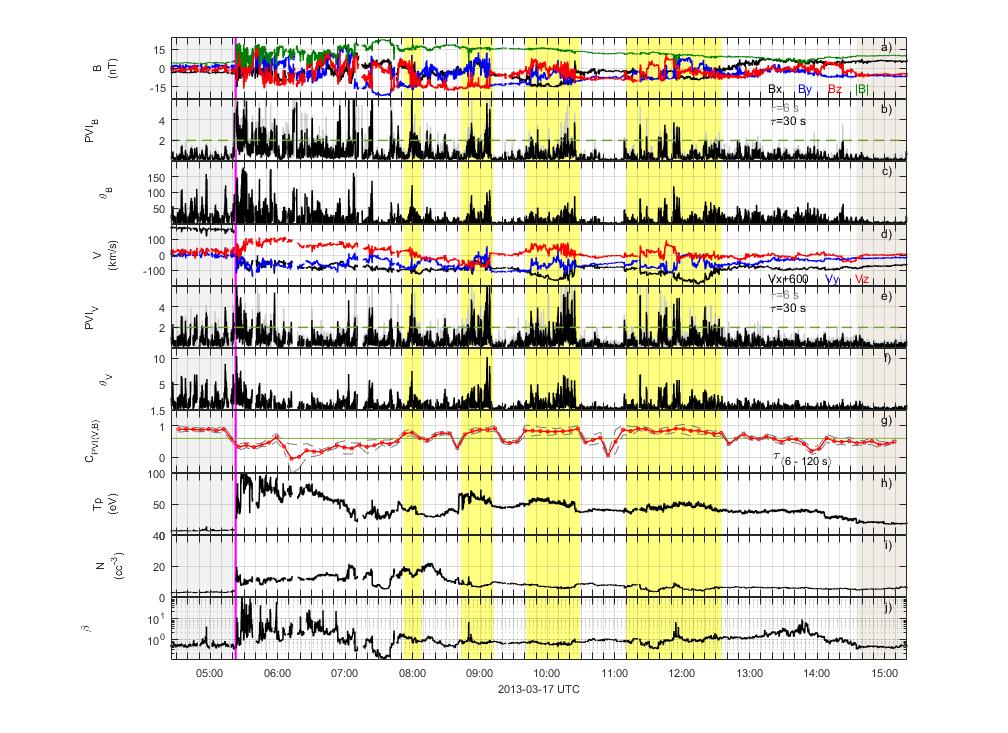}
 \caption{CME sheath observation by WIND spacecraft bounded by the gray shaded box marking the preceding solar wind and the beige box, marking the boundary layer at the magnetic cloud's leading edge. The vertical magenta line represents the associated interplanetary shock. The yellow boxes highlight the intervals with intermittent plasma heating. Panels: $a$) magnetic field components and intensity in GSE; $b$) PVI(B) for time delays $\tau=$30 s (in black) and $\tau=$6 s (in gray); $c$) magnetic field rotation for the same time delay for the same delays; $d$) ion bulk velocity (V$_z$ component shifted with 600 km/s); $e$) PVI(V) and $f$) velocity shear angle for the same $\tau$ values; $g$) correlation between PVI(B) and PVI(V), where the green horizontal line marks the threshold 0.6, the red dashed curve is the average correlation obtained for the range of time delays from 6 to 120 s (dashed gray curves); $h$) proton temperature; $i$) proton density; and $j$) proton plasma beta.}
\label{fig:fig7}
\end{figure}

Finally, we investigate an interval from an ICME event observed by WIND at L1 (Figure \ref{fig:fig7}). In this figure, we plot only the ICME sheath, which is the region confined between the IP shock (magenta vertical line) at 05:23 UTC on 17 Mar 2013 and the leading edge of the magnetic cloud at $\sim$14:35 UTC (defined by the temperature and plasma beta drop and the lack of strong fluctuations in all variables) on the same day. The preceding solar wind and the part of the magnetic cloud are shaded in gray and beige, respectively. The ICME was fast with shock speed $~$650 km/s and an average sheath speed 700 km/s. The shock Mach number M$_A$=6.1 and the magnetosonic number M$_S$=4.2, according to the \textit{IPshocks} database  (\url{http://ipshocks.fi/}).
The downstream geometry is quasi-parallel with an angle between the shock normal and the upstream magnetic field direction $\theta_{B_n}\sim$35$^{\circ}$. Overall, the entire sheath region is highly variable, with large fluctuations in the magnetic field often changing sign compared to the preceding solar wind (Fig. \ref{fig:fig7}, panel $a$). This is especially true in the part close to the IP shock, where there is also plasma compression (panel $i$), strong plasma heating (panel $h$), and high ion plasma beta (panel $j$). Similarly to the MSH cases, we calculate PVI(B) and PVI(V) of the ion bulk velocity, and magnetic and velocity shear angles from single point measurements using a time shift $\tau$=30 s (in black) in panels $b),c)$ and $e),f)$, respectively. The time scale is chosen to be as close as possible to the kinetic range and not to affect the accuracy of the PVI estimation by an insufficient number of points due to the low sampling frequency of the plasma data. To verify that the structures are current sheets (PVI$>$3) we also plot in gray the PVIs and shears from the smallest accessible scale $\tau$=6 s. The magnetic field makes several rotations from 0$^{\circ}$ to 180$^{\circ}$ behind the IP shock until 07:50 UTC (panel $b$). Some of these rotations are associated with high PVI(B) peaks (panel $c$), confirming the presence of current sheets. At the same time, there are no sharp gradients and rotations in the velocity field (panels $d, f$) in this interval, except for the PVI(V) peak accounting for the detection of the shock (panel $e$, the threshold is noted by the dashed green line). Looking deeper inside the sheath there are periods of smooth magnetic filed with no directional changes and shears adjacent to shorter periods (highlighted in yellow) characterized by peaks in both PVI(B) and PVI(V) (panels $b, e$), and strong magnetic shears (panel $c$). Although much smaller than the magnetic field ones, the velocity shears are significant relative to the quiet intervals with a smoother magnetic field, velocity, and temperature in between the yellow shaded boxes. Panel $g$) shows the correlation coefficient between PVI(B) and PVI(V) (red line with circles), obtained from running averages with $50\%$ overlapping windows of $\sim$ 260 points. The correlation coefficients are calculated as the average in scales ranging from 6 to 120 s (gray dashed lines), which fall into the higher frequency part of the MHD range close to the beginning of the kinetic range. Notably, there is a very strong correlation between PVI(B) and PVI(V) in the active regions (the horizontal green line marks the 0.6 threshold) which coincides with temperature enhancements. Interestingly, also plasma beta (panel $j$) is nearly constant in the interval 07:50-13:30 UTC comprising these active and quiet regions.
 
\section{Discussion} \label{sec:sum}

 The MSH interval studied here is relatively short (4 min long), however, represents well-calibrated data. The MSH geometry is quasi-parallel with several current sheets detected. Recently, observations in quasi-parallel MSH have revealed correlations between the electron heating and strong PVI(B) (\citealp[]{Chasapis15},\citealp{Chasapis17}) at abundant current sheets (\citealp[]{Retino07}, \citealp[]{Voros16}, \citealp[]{Yordanova20}). This particular MSH time interval has been the subject of many studies concerning different kinetic processes such as magnetic reconnection in thin current sheets (\citealp[]{Yordanova16}, \citealp[]{Voros17}), electron acceleration at a current sheet \citep{Erik16}, whistlers, and lower hybrid waves at a reconnection site \citep{Voros19a}, energy dissipation \citep{Voros19b}, and non-Maxwellian features of the ion velocity distribution functions \citep{Perri20}. This event is ideally suitable to test our hypothesis and methods on the replacement of MMS four-point high-resolution tetrahedron measurements with one-point proxies. The current sheets are represented by the sharp magnetic field gradients and rotations centered roughly, e. g. at ~00:22:40, 00:23:10, 00:23:30, 00:24:22, 00:25:03, 00:24:48, 00:25:50, 00:26:10 UTC, (Fig. \ref{fig:fig2}, panel $a$, $b$ and $c$). The curlometer $\mid\nabla\times\textbf{\textit{B}}\mid$ (panel $b$) was introduced as a proxy for the current density under the assumptions that the changes in the magnetic field between the spacecraft in the tetrahedron are time-independent and linear \citep{Dunlop02}. The correlation between the single point PVI(B) and four-point current density from the curlometer techniques has been established earlier in other magnetosheath observations (\citealp[]{Chasapis17}, \citealp[]{Yordanova20}). This is confirmed in our case as well, where we observe high correlation between the PVI(B) and $\mid\nabla\times\textbf{\textit{B}}\mid$ (panel $d$).

The current sheet intervals coincide with large fluctuations and changes in the orientation of the plasma flow (panel $e$). For example, in the interval containing the current sheet at 00:24:22, \citep[see their Fig.1]{Erik16}, have identified several jets using the criterion for detecting high pressure flows introduced by \citealp[]{Plaschke13}. This current sheet, seen as the strongest rotation in Fig.  \ref{fig:fig2}, panel $b$) and strong peaks in PVI(B) (panel $c$) is inside one of the jets. This jet can be recognized as the largest vorticity and PVI(V$_i$) as well (panel $f$ and $g$). The correlation between the vorticity and PVI(V$_i$) results from the burstiness of activity in both quantities (panel $h$). Given that the correspondence between one-point and four-point techniques (panels $d$ and $h$) shows good agreement in the detection of intermittent structures, we now correlate directly the PVI of the magnetic field and the ion velocity (panel $i$). The results show that the correlation is very high in the intervals containing the currents sheets accompanied at the same time with strong shear flows, in analogy with the prediction from numerical simulations \citep{Karimabadi13}.

Figure \ref{fig:fig4} represents the same quantities except that now the vorticity (panel $f$) and PVI(V$_e$) (panel $g$) are calculated from the electron speed (panel $e$). Thanks to the higher sampling rate we can see now more details in the single point PVI(V$_e$). In the beginning of the interval there are correlations (panels $d$, $h$, $i$) detected by the one-point parameters coming from smaller scales magnetic field and velocity gradients (panels $c$, $g$), which were not visible in the ion data (Fig. \ref{fig:fig2}). The same is true for the current sheet at 00:26:10, where correlations can be seen only in the electron data from the single point PVIs (Fig. \ref{fig:fig4}, panel $i$). This current sheet has been associated with signatures of a reconnection \citep{Yordanova16}, where no clear ion outflow but only electron jets have been found. There occurs also the strongest fluctuation in the velocity (Fig. \ref{fig:fig4}, panel $a$) and it is represented by the largest velocity gradient and shear (panels $f,g$). There is also a reverse case for the structure at 00:22:30-00:22:40 UTC, where there is no single point PVI(B,V$_e$) correlation in the electron data (panel $i$), but there is a correlation in the ion data (Fig. \ref{fig:fig2}, panel $i$). This indicates that the velocity and magnetic field gradients are possibly generated over larger ion scales. At this point, it is worth emphasizing the importance of the PVI conditioning in distinguishing the coherent structures from the noise. One should also consider whether the measurement resolution can resolve the scales which are under investigation, as well as to keep in mind in the case of multipoint measurement that the PVI detection would be limited to structures that are larger than the corresponding spatial separation of the spacecraft.
Apart from the above-mentioned examples, the intervals with correlations characterizing the rest of the regions of the currents sheet coincide with electron and ion data. This reflects the multi-scale nature of the intermittent coherent structures, in analogy to observations in the turbulent solar wind \citep{Greco16}, and that the current sheets are associated with regions of strong vortical flows \citep{Karimabadi13}. Vorticity and flow shears do not affect only the plasma dynamics  but can lead to additional instabilities driven by shear induced pressure anisotropy. A non-gyrotropic electron pressure tensor then represents the dominant non-ideal term in generalized Ohm’s law driving magnetic reconnection \citep{DelSarto16}.

We continue further discussing the application of the proposed method to the large KH vortices. The consecutive changes from high to low temperatures in Figure \ref{fig:fig5}, (panels $i, j$), are due to the spacecraft passing through different plasmas of LLBL and MSH origin. It is difficult to observe local peaks in the ion and electron temperature that are not associated with different regions themselves, however, near the LLBL boundaries, there are additional temperature elevations in the vicinity of the current sheets. These local enhancements in T$_i$ associated with the proxies and the curlometer and vorticity could be made visible when the large-scale variations were removed (red curve in panel $i$). 
Similarly, in \citep{Osman12} and \citep{Sorriso18} it was shown, that in order to identify correlations between turbulent structures and local heating a conditional average analysis is needed to allow the persistent but weak heating to emerge from the turbulent temperature fluctuations.

Inside the current sheet at ~11:02 UTC (yellow shaded), characterized by peaks in the PVI(B) (panel $b$) and magnetic shear (panels $c,d$), there is a plasma jet discerned by strong vorticity and large PVI(V$_i$) in (panel $g,h$). At the same interval, both ion and electron temperatures are elevated. This jet was recognized previously as a result of magnetic reconnection \citep{Eriksson16}, where also two of the MMS spacecraft crossed the electron diffusion region and observed significant electron parallel heating \citep{Eriksson17}. One can see similar features of enhanced ion and electron temperature, and high gradients and vorticity in other current sheets (yellow shaded). In fact, this more than $2$-hour KH event has been recognized as the first observation of reconnection exhausts occurring in more than a half of the cases of the current sheets associated with the KH waves \citep{Eriksson16}. Therefore, we can relate directly some of the observed plasma heating with the energy dissipation due to magnetic reconnection.

With the events of MSH turbulence and large KH vortices at the magnetospheric flanks, we have shown that single-point PVIs represent well the detected gradients and shears by the multi-spacecraft proxies. While in the MSH turbulence there can be more than one competing kinetic processes responsible for the energy conversion  that can be accounted for the plasma heating \citep{Voros19b}, in the KH vortices with this method we could connect individual current sheets with plasma heating.

Despite the recent advances in collisionless plasma heating problem (\citealp[]{Tong19}, \citealp[]{Halekas20}, \citealp[]{Lopez20}, \citealp[]{Zhao20}), it is still unclear how heating is generated in larger plasma volumes with numerous intermittent current sheets and shear flows. Although the comprehension of individual heating events is crucial, the volume integrated heating, even if it involves specific field and plasma structures only, might offer more understanding of energy conversions in real large-scale systems such as the solar corona or the solar wind. For this purpose, we apply the method to the turbulent ICME sheath, which is part of a large interplanetary transient. 
It is very clear from Figure \ref{fig:fig7}, that $\sim$ 2 hours downstream of the IP shock (magenta vertical line), the temperature enhancements occur intermittently in blobs of a duration of tens of minutes (in yellow) corresponding to the range $12000-35000$ ion inertial lengths. In terms of the introduced proxies, these blobs contain the current sheets (enhanced PVI(B) and rotation angle), the vortical sheared flows (enhanced PVI(V) and rotation angle), and correlations between them that are over the threshold. In other words, we refer to such blobs as groups of current sheets associated with vortices and high plasma temperature. 
Obviously, the heating mechanisms in the vicinity of the IP shock manifested in low correlations between PVI(V, B) (panel $j$) but high T$_i$ (panel $h$) are different. 
We can also speculate that the proposed method works rather well in this event, because of the non-fluctuating low beta plasma. 
In ICME sheaths with strongly fluctuating plasma beta, it would be more difficult to observe the outlined correlations. 
Investigations into this direction are postponed to further studies.

\section{Conclusions} \label{sec:con}
In this paper, we introduced a new pathway for plasma heating in intermittent blobs not considered before in space plasma experiments. Although there is no exact theory for coherent structure formation near ion scales, there exist simulation and experimental results that turbulence can generate such structures. Since the blobs are much larger than the ion inertial length (100$R_E$ or larger) the turbulent processes generating these intermittent structures are of fluid scale. The one-point proxies replacing the four-point $\mid\nabla\times\textbf{\textit{B}}\mid$ ($= \textbf{j}$) and $\mid\nabla\times\textbf{\textit{V}}\mid$ (vorticity) are introduced from necessity to understand better turbulent heating in the solar wind. Because of the above reasons, the proxies are not based on rigorous mathematical relations but represent approximations and generalizations corresponding to previous simulation and experimental results. The proxies seemed to be useful in describing heating in larger blobs in the sheath region of an ICME, where plasma beta was low without strong fluctuations. It is known from previous studies \citep{Phan10} that when plasma beta is changing strongly across a reconnecting current sheet, reconnection is stopped or the current sheet is destroyed. This suggests that further statistical studies based on the proposed proxies will be needed to investigate the parameter space validating this approach under different conditions. It is also a challenge for further analysis to identify the turbulent multi-scale structures which are responsible for the generation or absence of blobs.

\typeout{}

\acknowledgments
EY, LSV and APD have been supported by Swedish Contingencies Agency, grant 2016-2102. ZV has been supported by the Austrian FWF under contract P28764-N27. EK acknowledges the ERC under the European Union's Horizon 2020 Research and Innovation Programme Project 724391 (SolMAG), Academy of Finland Project 310445 (SMASH). European Union’s Horizon 2020 research and innovation programme under grant agreement No 101004159 (SERPENTINE). EK contribution has been achieved under the framework of the Finnish Centre of Excellence in Research of Sustainable Space (FORESAIL; Academy of Finland grant number 312390), which she gratefully acknowledges.


\end{document}